\documentclass{article}

\usepackage{arxiv}

\usepackage[utf8]{inputenc} 
\usepackage[T1]{fontenc}    
\usepackage[hidelinks]{hyperref}       
\usepackage{url}            
\usepackage{booktabs}       
\usepackage{amsfonts}       
\usepackage{nicefrac}       
\usepackage{microtype}      
\usepackage{lipsum}		
\usepackage{graphicx}
\usepackage{natbib}
\usepackage{doi}

\usepackage{amsmath}
\usepackage{amsthm}
\usepackage{amssymb}
\usepackage{cleveref}
\usepackage{graphicx}
\usepackage{geometry}
\usepackage{xcolor}
\usepackage{rotating}
\usepackage{comment}

\usepackage{setspace}

\let\proglang\textsf

\title{Similarity network aggregation for the analysis of glacier ecosystems}

\author{ \href{https://orcid.org/0000-0002-7148-1468}{\includegraphics[scale=0.06]{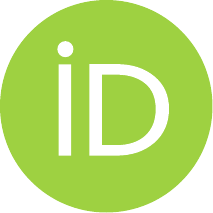}\hspace{1mm}Roberto Ambrosini}\\
	Dipartimento di Scienze e Politiche Ambientali\\
	Università degli Studi di Milano\\
	Via Celoria, 26 - Corpo C, Milano, 20133, Italy \\
	\texttt{roberto.ambrosini@unimi.it} \\
	\And
	\href{https://orcid.org/0000-0001-6107-9611}{\includegraphics[scale=0.06]{orcid.pdf}\hspace{1mm}Federica Baccini}\thanks{Corresponding author.}  \\
	Dipartimento di Ingegneria informatica, automatica e \\gestionale  ``Antonio Ruberti" \\
	Università degli Studi di Roma ``La Sapienza"\\
	Via Ariosto, 25, Roma, 00185, Italy \\
	\texttt{baccini@diag.uniroma1.it} \\
        \And
	\href{https://orcid.org/0000-0002-7210-4541}{\includegraphics[scale=0.06]{orcid.pdf}\hspace{1mm}Lucio Barabesi} \\
	Dipartimento di Economia Politica e Statistica, \\
	Università degli Studi di Siena\\
	Piazza San Francesco, 7, Siena, 53100, Italy \\
	\texttt{lucio.barabesi@unisi.it} \\
}

\date{}

\renewcommand{\headeright}{}
\renewcommand{\undertitle}{}
\renewcommand{\shorttitle}{Similarity network aggregation for the analysis of glacier ecosystems}

\begin{document}

\maketitle

\begin{abstract}
The synthesis of information deriving from complex networks is a topic receiving increasing relevance in ecology and environmental sciences. In particular, the aggregation of multilayer networks, i.e.\ network structures formed by multiple interacting networks (the layers), constitutes a fast-growing field. In several environmental applications, the layers of a multilayer network are modelled as a collection of similarity matrices describing how similar pairs of biological entities are, based on different types of features (e.g.\ biological traits). The present paper first discusses two main techniques for combining the multi-layered information into a single network (the so-called monoplex), i.e.\ Similarity Network Fusion (SNF) and Similarity Matrix Average (SMA). Then, the effectiveness of the two methods is tested on a real-world dataset of the relative abundance of microbial species in the ecosystems of nine glaciers (four glaciers in the Alps and five in the Andes). A preliminary clustering analysis on the monoplexes obtained with different methods shows the emergence of a tightly connected community formed by species that are typical of cryoconite holes worldwide. Moreover, the weights assigned to different layers by the SMA algorithm suggest that two large South American glaciers (Exploradores and Perito Moreno) are structurally different from the smaller glaciers in both Europe and South America. Overall, these results highlight the importance of integration methods in the discovery of the underlying organizational structure of biological entities in multilayer ecological networks.	
\end{abstract}


\section{Introduction}\label{sec::intro}

The analysis of complex networks is a rapidly expanding area in data science, since many real-world structures can be explored by assessing the connections and the interplay between their parts \citep{newman2018networks,estrada2012structure,barabasi2013network}. In the field of ecology, \citet{mittelbach2019} is an authoritative source for appraising the growing use and application of complex networks. Environmental and ecological networks aim at modelling the interactions among a set of biological and abiotic entities, which are represented as nodes of a graph. Interactions characterize living systems, because organisms necessarily interact with each other and with the environment. The reciprocal actions between ecological populations are the basis of community ecology and are central to shape biodiversity. As an example of an ecological network, the different plant species in a certain ecosystem could be interpreted as the graph nodes, while pollination by the same pollinator species represents the link between them. 

Recent advances in network science have pointed out that standard networks, i.e.\ networks constituted by pairwise interactions among a set of nodes, are often not sufficient to describe the full complexity of ecological interactions \citep{bianconi2018multilayer,de2022multilayer}. In order to implement a more realistic depiction of an ecological system, a key ingredient consists in considering different types of interactions between organisms \citep{hutchinson2019,pilosof2017}. Indeed, by continuing the previously considered example, different plant species in an ecosystem are likely to be pollinated by multiple pollinator species, instead of a single one. In this setting, multilayer networks \citep{kivela2014multilayer,bianconi2018multilayer,de2022multilayer} are an efficient way to represent systems where multiple types of interactions (i.e.\ the layers) occur between the nodes. More explicitly, each layer of a multilayer network constitutes a graph, where the edges are representative of a given type of interaction. The nodes may be partially different in different layers, and even interlayer edges may be present. \citet{hutchinson2019} provide many real examples of multilayer networks in ecology, with special emphasis on the interactions among species.

Different families of multilayer networks exist, each aimed at modelling different settings. A frequently-adopted type of multilayer network is the so-called `multiplex' \citep{bianconi2018multilayer,de2022multilayer}. A multiplex network has the same set of nodes on each layer (the node `replicas'), and each layer represents a different relation type. The present work focuses on multiplex networks and, in particular, on techniques for aggregating the multi-layered information into a single graph, i.e.\ the so-called monoplex. The resulting aggregated network may be central to evaluate the strong connections between the different nodes and -- eventually -- to implement a cluster analysis for searching cohesive groups in the original multiplex network. In addition, an appropriate analysis of the correlation between the monoplex and the single layers may be helpful to detect the global community structure.

We emphasize that a multiplex is often used to model similarities between pairs of entities (the nodes), where the similarity is computed based on several feature types. This defines a multiplex similarity network consisting of a collection of adjacency similarity matrices, whose dimension is given by the number of nodes (see e.g.\ \cite{baccini2022average}). In the framework of multiplex similarity networks, we discuss two main techniques for integrating the similarity networks of the multiplex in order to obtain a single monoplex with suitable properties. The first technique is Similarity Network Fusion, the widely-used approach initially suggested by \citet{wang2012unsupervised} and \citet{wang2014similarity}. Then, the more recent approach introduced by \citet{baccini2022average}, namely SimilarityMatrix Average, is revised and discussed. The latter approach is designed to ``average" a collection of similarity matrices by exploiting advances in the theory of barycenters in appropriate matrix spaces. 

The effectiveness of the two methods for the study of ecological networks is tested on a real-world dataset describing the abundance of microbial species in glacier ecosystems. Glaciers host active ecological communities and have been recognized as ecosystems in their own right, where the dominant life forms are microbes, especially bacteria \citep{hodson2008glacial,anesio2012glaciers}. Most ecological studies on glaciers have focused on cryoconite holes, i.e.\ small ponds that form on the glacier surface with a tiny layer of debris at the bottom, and are considered as the most biodiverse ecosystems in these harsh environments \citep{cook2016cryoconite,ambrosini2017diversity}. The scope of this analysis is twofold. First, we aim at giving insights into the similarities between microbial species based on their abundance in different glaciers. Subsequently, on the basis of the similarity between the analysed microbial species, we assess how much distinct glaciers are similar in terms of the relative abundance of microbial species.

The paper is organized as follows. Section \ref{sec::methods} formalizes the problem of integrating the information on a multiplex similarity network; the same section also contains the description and the analysis of the techniques adopted for the similarity network aggregation procedure. Section \ref{sec::3} presents the application of the considered methods to the dataset registering the relative abundance of microbial species in cryoconiye holes of glaciers worldwide. Finally, some conclusions are drawn in Section \ref{sec::conclusion}.

\section{Integration of multiplex similarity networks}\label{sec::methods}

In this section the problem of integrating the multiplex similarity network into the monoplex is described, by assuming two main possible geneses for the multiplex. In both geneses, the multiplex is modelled as a set of similarity matrices. 

A first framework where a multiplex similarity network can originate is the case in which $m$ measurements referring to a fixed set of entities are available. 
 
For each measurement, it is possible to define a graph (the ``similarity network") where the vertices are the items, and the edges are weighted with a suitable similarity score between items. The similarity can be computed via suitable similarity measures, many choices of which are available (see \cite{newman2018networks,eck2009normalize,debello2021}). The intuitive idea behind the use of similarity measures is that a high similarity score is associated to pairs of objects having a similar profile based on some measurement. Then, in the resulting similarity network, such objects will be connected by a tighter edge. 

If $m$ measurements are available, $m$ similarity networks (the ``layers"), each one formed by the same set of nodes, are originated, and the resulting structure is referred to as ``multiplex similarity network". Each layer of the multiplex is characterized by a similarity network whose weighted adjacency matrix is the corresponding similarity matrix. As a real-world example in the ecological setting, the items could represent $n$ species in an ecosystem, and the measurements could denote $m$ distinct functional traits. Similarities between species can then be computed based on each trait, thus producing a multiplex similarity network where each layer represents a specific trait. 

Formally, a multiplex similarity network is represented by means of $m$ similarity matrices of order $(n\times n)$, namely $\mathbf{S}_1,\dots,\mathbf{S}_m$. 

A formalization of the genesis of such multiplex similarity network is now necessary. A generic entity, or object, is described, for the $l$-th measurement, as a vector of $n$ observations, say $(x_{1l},\dots,x_{nl})$, for $l=1,\dots,m$. In general, the $n$ observations could be also multivariate and can assume values in a given Euclidean space. If the $l$-th measurement is discrete or continuous, one option to compute similarities between observations is to consider a transform of the Euclidean distance as a measure of similarity. Thus, a possible choice for the elements of the similarity matrix $\mathbf{S}_l = (s_{ij}^{(l)})$ is given by
\begin{equation}\label{eq::wandsim}
    s_{ij}^{(l)} = \exp{\left(-\frac{d_\mathrm{E}(x_{il},x_{jl})^2}{\sigma}\right)},
\end{equation}
\noindent
where $\sigma>0$ is a scaling parameter to be fixed, and $d_\mathrm{E}$ is the appropriate Euclidean distance (see e.g. \cite{wang2012unsupervised}). Alternatively, if the $l$-th measurement $x_{il}$ assumes values in $\{0,1\}$ for $i=1,\dots,n$, the similarity matrix $\mathbf{S}_l=(s_{ij}^{(l)})$ could be chosen as follows
\begin{equation}\label{eq::presencesim}
    s_{ij}^{(l)} = \mathbf{1}_{\{x_{il} = x_{jl}\}},
\end{equation}
\noindent
where $\mathbf{1}_A$ represents the indicator function of a set $A$ (see e.g.\ \cite{debello2021}). When \eqref{eq::presencesim} is adopted, the similarity matrix $\mathbf{S}_l$ is a joint presence-absence matrix. When \eqref{eq::wandsim} or \eqref{eq::presencesim} are selected, $s_{ij}^{(l)}$ turns out to be a normalized similarity measure in $[0,1]$ and $\mathbf{S}_l$ is symmetric.

A second framework where a multiplex structure arises concerns the modelling of relations among a fixed set of $n$ groups described by the membership to such groups of $m$ different sets of elements. This situation can be modelled by using a multilayer network characterized by the following two properties: (i) each layer is a bipartite network, and (ii) one vertex set of the bipartition is common to all the layers (the set of $n$ groups), while the other one can change between layers (each of the $m$ different sets of elements). As a real-world example in the ecological field, one can consider $n$ groups of herbivore species that interact with some plant species, in $m$ distinct ecosystems. A task of interest could be the analysis of the relations among the $n$ herbivore species, taking into account all the distinct ecosystems. This situation can be modelled as a multilayer network structured as above, where each layer represents an ecosystem, and contains a bipartite network formed by the set of herbivore species (common to all the layers), and the set of plant species they graze in each ecosystem. 

If the above multilayer structure is assumed, a multiplex network could be useful to explore the relations within the set of nodes that are common to all the layers. The multiplex can be obtained by considering the ``one-mode projection" of each layer onto the common subset of nodes (the groups), which can in turn be used to compute suitable similarities among the set of groups. The one-mode projection is a particular network that can be computed starting from a bipartite network (see e.g.\ \cite{newman2018networks}). More specifically, let $\mathbf{B}_l=(b_{ij}^{(l)})$ denote the incidence matrix of order $(p_l\times n)$ corresponding to the bipartite network on the $l$-th layer, whose generic elements are defined as follows
\begin{equation*}
b_{ij}^{(l)} = 
    \begin{cases}
        1 & \text{if item } i\text{ in the } l\text{-th layer belongs to group }j \\
        0 & \text{otherwise. }
    \end{cases}
\end{equation*}
Coherently with the previously-used terminology, the $l$-th bipartite network can be interpreted as describing the membership of $p_l$ items to $n$ groups. 
\noindent
The one-mode projection on the set of $l$ items can then be defined as $\mathbf{G}_l=(g_{ij}^{(l)})=\mathbf{B}_l^\mathrm{T}\mathbf{B}_l$. In particular, $\mathbf{G}_l$ is a symmetric matrix of order $(n\times n)$, whose generic element $g_{ij}$ represents the number of items of the $l$-th layer which belong to both the $i$-th and $j$-th group. 

Interestingly, as stated above, the one-mode projection $\mathbf{G}_l$ may be used to compute a similarity matrix $\mathbf{S}_l$. The cosine and the Jaccard similarity are two popular examples of similarity measures that can be computed starting from the one-mode projection $\mathbf{G}_l$ (see e.g.\ \cite{newman2018networks}, Sec. 7.6). Indeed, the Jaccard coefficient is formally given by
\begin{equation}\label{eq::jaccardsim}
    s_{ij}^{(l)} := J_{ij}^{(l)}
    = \frac{g_{ij}^{(l)}}{g_{ii}^{(l)}+g_{jj}^{(l)}-g_{ij}^{(l)}},
\end{equation}
\noindent
while the cosine similarity turns out to be
\begin{equation}\label{eq::cossimilarity}
    s_{ij}^{(l)} := C_{ij}^{(l)}
    = \frac{g_{ij}^{(l)}}{(g_{ii}^{(l)}g_{jj}^{(l)})^{1/2}}
\end{equation}
\noindent
(for more details, see \cite{baccini2022average}). Finally, by assuming \eqref{eq::jaccardsim} and \eqref{eq::cossimilarity}, $s_{ij}^{(l)}$ is a normalized similarity measure in $[0,1]$ and $\mathbf{S}_l$ is symmetric, since $\mathbf{G}_l$ is symmetric.

\subsection{Similarity network fusion}\label{subsec::snf}

Similarity Network Fusion (SNF) is an unsupervised technique first introduced by \citet{wang2014similarity} with the purpose of integrating multilayered data. Specifically, given a multiplex similarity network, SNF iteratively updates the similarity matrix corresponding to each layer in order to obtain a unique similarity network, which is an ``averaged" representation of the single layers. The technique has been successfully used in many different frameworks. For example, in genetics and epigenetics, \citet{wang2014similarity} and \citet{baccini2022graph} finalize the methodology to the aggregation of different types of genome data; \citet{baccini2022similarity} consider instead an application to scientometrics by integrating a multiplex similarity network built on different scholar interactions between scientific journals.

SNF is built upon an iterative procedure named Cross Diffusion Process (CDP) suggested by \citet{wang2012unsupervised}. CDP aims at enhancing two types of edges in the desired monoplex, i.e.\ strong links which are present in some layers, and links which are shared by all the layers. Let us assume that $\mathbf{P}_l = (p_{ij}^{(l)})$ denotes the normalized matrix obtained by $\mathbf{S}_l$ as follows
\begin{equation*}
    p_{ij}^{(l)} = \frac{s_{ij}^{(l)}}{\sum_{k,h=1}^n s_{kh}^{(l)}}.
\end{equation*}
In addition, for the $l$-th layer, a further ``locally" normalized matrix is introduced, namely $\mathbf{Q}_l=(q_{ij}^{(l)})$, where 
\begin{equation*}\label{eq::locsim}
    q_{ij}^{(l)} = \begin{cases}
        \frac{s_{ij}^{(l)}}{\sum_{h\in N_i} s_{ih}^{(l)}} & j\in N_i \\
        0 & \text{otherwise},
    \end{cases}
\end{equation*}
\noindent
and $N_i$ represents the index set of the $k$-th nearest neighbours for the $i$-th node, computed via the $k$-Nearest Neighbours ($k$-NN) algorithm. We recall that the parameter $k$ must be specified in advance, and it is fixed throughout the CDP procedure. For more details about the $k$-NN algorithm, see e.g.\ \citet{hastie2009elements}. The introduction of $\mathbf{Q}_l$ aims at eliminating the mutual influence between distant nodes in the $l$-th layer. Indeed, the edge weight between non-neighbouring nodes in the $l$-th layer is null in $\mathbf{Q}_l$.

The matrices $\mathbf{P}_1,\ldots,\mathbf{P}_m$ are called ``initial status" matrices, since they constitute the first step of the iterative procedure CDP. Let $\mathbf{P}_{l,t}$ denote the $l$-th matrix computed at the $t$-th iteration of the CDP. For the $l$-th layer, the updating rule is given by
\begin{equation*}\label{eq::updateruleSNF}
    \mathbf{P}_{l,t+1} = \mathbf{Q}_l \left(\frac{1}{m-1}
    \sum_{h\ne l = 1}^m\mathbf{P}_{h,t} \right) \mathbf{Q}_l^\mathrm{T},
\end{equation*}
\noindent
for $t=0,1,\dots$, and where $\mathbf{P}_{l,0}=\mathbf{P}_l$ for $l=1,\dots,m$.

\citet{wang2012unsupervised} show that CDP converges, in the sense that $\|\mathbf{P}_{l,t}-\mathbf{P}_{l,\infty}\|_\mathrm{F}\rightarrow 0$ as $t\rightarrow\infty$, where $\mathbf{P}_{l,\infty}$ is a suitable limit matrix, while $\|\mathbf{X}\|_{\mathrm{F}}=\mathrm{tr}(\mathbf{X}^{\mathrm{T}}\mathbf{X})^{1/2}$ represents the usual Frobenius norm of a matrix $\mathbf{X}$. Moreover, let us assume that there exists $t=T$ for which convergence is reached according to the accuracy criterion $\|\mathbf{P}_{l,T+1}-\mathbf{P}_{l,T}\|_\mathrm{F}<\varepsilon$ for each $l=1,\ldots,m$ and for a pre-fixed $\varepsilon>0$. Thus, the last step of SNF consists in averaging the matrices $\mathbf{P}_{1,T},\dots,\mathbf{P}_{m,T}$, i.e.\
\begin{equation}\label{eq::SNFbar}
    \mathbf{P}_{+T} = \frac{1}{m}\sum_{l=1}^m \mathbf{P}_{l,T}.
\end{equation}
Finally, the matrix \eqref{eq::SNFbar} is taken as the adjacency matrix of the desired monoplex. A similarity matrix, say $\mathbf S_\mathrm{SNF}$, is then suitably obtained by re-weighting $\mathbf{P}_{+T}$. 

Besides the enforcement of strong links, SNF also boosts weaker links that are shared by many layers, in the sense that densely connected groups of nodes are maintained and strengthened with tighter links in the monoplex, since they potentially represent strong relationships among groups of elements. However, some drawbacks in the use of SNF should be highlighted. A first observation is that the choice of the parameter $k$ in the $k$-NN algorithm is rather arbitrary, and different values of $k$ might result in significantly different monoplex structures. Second, each layer equally contributes to the structure of the monoplex defined by \eqref{eq::SNFbar}, i.e.\ SNF implicitly assumes that all the layers have the same relevance in determining the overall structure of the relations. 

\subsection{Similarity matrix average}\label{subsec::sma}

In the present section, we analyse the Similarity Matrix Average (SMA) method suggested by \citet{baccini2022average} for computing a monoplex network starting from a multiplex similarity network. SMA relies on some mathematical properties of similarity matrices, and exploits some recent advances in the theory of barycenters of objects lying in abstract spaces, with the ultimate goal of computing a matrix average \citep{bhatia2019procrustes,alvarez2016fixed}. 

Although results concerning matrix barycenters are usually considered in the space of positive definite matrices \citep{bhatia2019procrustes}, similarity matrices are more naturally handled in the space of completely positive matrices. Formally, the space $\mathcal{C}\mathcal{P}_n$ of completely positive matrices is defined as
\begin{equation*}
    \mathcal{C}\mathcal{P}_n = \{\mathbf{X}\in\mathbb{R}^{n\times n}: \mathbf{X}=\mathbf{Y}^{\mathrm{T}}\mathbf{Y},\exists\mathbf{Y}\in\mathbb{R}^{p\times n},
    \mathbf{Y}\geq\mathbf{0}\},
\end{equation*}
where the notation $\mathbf{X}\geq\mathbf{0}$ indicates that the matrix $\mathbf{X}$ has nonnegative elements. For details on the properties of this class of matrices, we refer the reader to the monographs by \citet{berman2003completely,shaked2021copositive}. If $\mathcal{P}_n$ represents the space of symmetric positive semidefinite matrices, a necessary condition for a matrix $\mathbf{X}$ to be completely positive is that $\mathbf{X}\in\mathcal{P}_n$ and $\mathbf{X}\geq\mathbf{0}$, even if -- contrary to intuition -- this condition is not generally sufficient. Hence, $\mathcal{C}\mathcal{P}_n$ is a proper subset of $\mathcal{P}_n$ and, in addition, $\mathcal{C}\mathcal{P}_n$ is a closed convex cone in $\mathcal{P}_n$ \citep{berman2003completely}. 

\citet{baccini2022average} show that several commonly-adopted similarity matrices are completely positive. Noticeably, the similarity matrices obtained by using the Jaccard index \eqref{eq::jaccardsim} and the cosine similarity \eqref{eq::cossimilarity} belong to the space $\mathcal{C}\mathcal{P}_n$. The fact that many similarity matrices belong to $\mathcal{C}\mathcal{P}_n$ is of central importance for SMA. Indeed, in SMA the problem of obtaining a monoplex starting from a multiplex similarity network is formulated as a minimization problem carried out in the space of completely positive matrices. Formally, the adjacency matrix of the monoplex can be computed as the barycenter $\mathbf{S}_+$ of the $m$ similarity matrices $\mathbf{S}_1,\dots,\mathbf{S}_m$ representing the layers, by solving the following minimization problem
\begin{equation}\label{eq::riembar}
    \mathbf{S}_+ = \mathrm{arg}\min_{\mathbf{X}\in\mathcal{C}\mathcal{P}_n}
    \sum_{l=1}^m w_l d(\mathbf{X},\mathbf{S}_l)^2,
\end{equation}
\noindent
where $d:\mathcal{C}\mathcal{P}_n\times\mathcal{C}\mathcal{P}_n\rightarrow\mathbb{R}^+$ is a metric on $\mathcal{C}\mathcal{P}_n$, and the weights $\mathbf{w}=(w_1,\ldots,w_m)^\mathrm{T}$ are such that $w_l\ge 0$ and $\sum_{l=1}^m w_l=1$. In a general framework, $\mathbf{S}_+$ provides the so-called Fréchet mean (see e.g.\ \cite{bacak2014computing}). The Fréchet mean might not be suitable in an arbitrary metric space, though being highly appropriate in a geodesic metric space of non-positive curvature, i.e.\ a Hadamard space such as $\mathcal{C}\mathcal{P}_n$. The existence and uniqueness of the minimizer $\mathbf{S}_+$ in a Hadamard space are assured by Theorem 2.4 by \citet{bacak2014computing}. Obviously, the optimization problem in \eqref{eq::riembar} also depends on the choice of the metric $d$ and of the weights $\mathbf{w}$. 

Concerning the metric choice, the cases of the Frobenius, Riemannian and Wasserstein metrics are discussed in \citet{baccini2022similarity}. If the classical Frobenius metric is adopted in \eqref{eq::riembar}, i.e.\
\begin{equation*}\label{def:frobmetric}
    d_{\text{F}}(\mathbf{X},\mathbf{Y}) = \|\mathbf{X}-\mathbf{Y}\|_{\mathrm{F}}
    = \mathrm{tr}((\mathbf{X}-\mathbf{Y})^{\mathrm{T}}(\mathbf{X}-\mathbf{Y}))^{1/2},
\end{equation*}
we obtain the weighted matrix average suggested by \citet{abdi2005distatis}, i.e. 
\begin{equation}\label{eq::frobeniusbar}
    \mathbf{S}_{+\mathrm{F}}=\sum_{l=1}^m w_l\mathbf{S}_l.
\end{equation}
\citet{baccini2022average} emphasize that $\mathbf{S}_{+\mathrm{F}}\in\mathcal{C}\mathcal{P}_n$, since \eqref{eq::frobeniusbar} is a weighted sum of completely positive matrices with positive weights. However, this proposal may involve a ``swelling effect'' as a drawback, meaning that $\mathbf{S}_{+\mathrm{F}}$ may show an increase in the determinant with respect to the single addends of the matrix average \eqref{eq::frobeniusbar}. As a result, the final monoplex might not be highly representative of the information contained in the multiplex. 

An alternative choice for $d$ in \eqref{eq::riembar} is the Riemannian metric, i.e.
\begin{equation*}\label{def:riemandist}
    \begin{aligned}
        d_{\mathrm{R}}(\mathbf{X},\mathbf{Y}) &
        = \|\log(\mathbf{X}^{-1/2}\mathbf{YX}^{-1/2})\|_\mathrm{F}\\ &
        = \mathrm{tr}(\log(\mathbf{X}^{-1/2}\mathbf{YX}^{-1/2})^\mathrm{T}
        \log(\mathbf{X}^{-1/2}\mathbf{YX}^{-1/2}))^{1/2}.
    \end{aligned}
\end{equation*}
\citet{bhatia2009positive} (Chapter 6, Theorem 6.1.6) remarks that $d_{\mathrm{R}}$ naturally arises in the framework of Riemannian geometry (for further details, see also \cite{bhatia2019procrustes}, and references therein). Moreover, $d_{\mathrm{R}}$ may be considered as the matrix version of the Fisher-Rao metric for probability laws (see in turn Chapter 6 in \cite{bhatia2009positive}). For this metric choice, the matrix given by \eqref{eq::riembar}, say $\mathbf{S}_{+\mathrm{R}}$, exists and is the unique solution of the nonlinear matrix equation in $\mathbf{X}$ given by 
\begin{equation}\label{eq:nonlinear}
    \sum_{l=1}^{m}w_l\log(\mathbf{X}^{1/2}\mathbf{S}_l^{-1}\mathbf{X}^{1/2})=\mathbf{0},
\end{equation}
even if it is expressible in closed form solely for $m=2$ (see e.g.\ \cite{lim2014weighteda,lim2014weighted}). For $m>2$, \citet{lim2014weighteda} introduce an iterative procedure which produces the solution $\mathbf{S}_{+\mathrm{R}}$ of \eqref{eq:nonlinear}. In addition, $\mathbf{S}_{+\mathrm{R}}$ may be less prone than $\mathbf{S}_{+\text{F}}$ to the swelling effect \citep{lim2014weighteda,lim2014weighted}.

A final proposal for $d$ in \eqref{eq::riembar} is given by the Wasserstein metric, i.e.
\begin{equation*}\label{def:wdist}
    d_{\text{W}}(\mathbf{X},\mathbf{Y}) = 
    \mathrm{tr}(\mathbf{X}+\mathbf{Y}-2(\mathbf{X}^{1/2}\mathbf{YX}^{1/2})^{1/2})^{1/2},
\end{equation*}
\noindent
which is of great importance in the theory of optimal transport and quantum information (for more details, see e.g.\ \cite{bhatia2019bures}). \citet{bhatia2019bures} emphasize that $d_{\mathrm{W}}$ displays many interesting features and, among others, it corresponds to a metric in Riemannian geometry. For this metric selection, \citet{alvarez2016fixed} show that the matrix given by \eqref{eq::riembar}, say $\mathbf{S}_{+\mathrm{W}}$, is the unique solution of the following nonlinear matrix equation in $\mathbf{X}$ 
\begin{equation}\label{eq:nonlinear2}
    \mathbf{X}=\sum_{l=1}^m w_l(\mathbf{X}^{1/2}\mathbf{S}_l\mathbf{X}^{1/2})^{1/2}.
\end{equation}
\noindent
\citet{baccini2022average} also remark that $\mathbf{S}_{+\mathrm{W}}\in\mathcal{C}\mathcal{P}_n$ (see also \cite{bhatia2019bures}). The solution of the nonlinear matrix equation \eqref{eq:nonlinear2} is known in a closed form only in the case $m=2$ (see e.g.\ \cite{baccini2022average}). For $m>2$, \citet{alvarez2016fixed} introduce a fixed-point algorithm which generates as a limiting matrix the solution $\mathbf{S}_{+\mathrm{W}}$ of \eqref{eq:nonlinear2}. It is worth noticing that $(\mathbf{S}_{+\mathrm{F}}-\mathbf{S}_{+\mathrm{W}})$ is positive semidefinite based on Theorem 9 by \citet{bhatia2009positive}. This property implies that the swelling effect is mitigated, unlike in the case of the Frobenius metric.
 
In order to introduce a possible choice of the elements of the weight vector $\mathbf{w}$, a measure of ``closeness" between each pair of the $m$ similarity matrices $\mathbf{S}_1,\ldots,\mathbf{S}_m$ is needed. A suitable measure is given by the RV coefficient proposed by \citet{robert1976unifying}. The matrix $\mathbf{R} =(r_{ij})$ of order $(m\times m)$ is then considered, where $r_{ij}$ represents the RV coefficient between $\mathbf{S}_i$ and $\mathbf{S}_j$, i.e.\
\begin{equation*}
    r_{ij}=\frac{\langle\mathbf{S}_i,\mathbf{S}_j\rangle_\mathrm{F}}
    {\|\mathbf{S}_i\|_\mathrm{F}\|\mathbf{S}_j\|_{\mathrm{F}}},
\end{equation*}
\noindent
where $\langle\mathbf{X},\mathbf{Y}\rangle_\mathrm{F}=\mathrm{tr}(\mathbf{X}^{\mathrm{T}}\mathbf{Y})$ represents the Frobenius inner product for two matrices $\mathbf{X}$ and $\mathbf{Y}$ of the same order. It holds that $r_{ij}\in[0,1]$, and the closeness between $\mathbf{S}_i$ and $\mathbf{S}_j$ increases as $r_{ij}$ approaches one.

When $\mathbf{S}_{+\mathrm{F}}$ is adopted, \citet{abdi2005distatis} suggest to consider the eigendecomposition of $\mathbf{R}$, i.e.\ $\mathbf{R}=\mathbf{Q}\mathbf{\Lambda}\mathbf{Q}^{\mathrm{T}}$, where $\mathbf{Q}=(\mathbf{q}_1,\ldots,\mathbf{q}_m)$ is the orthogonal matrix whose columns are the eigenvectors of $\mathbf{R}$, and $\mathbf{\Lambda}$ is the diagonal matrix whose entries are the positive eigenvalues of $\mathbf{R}$ considered in nonincreasing order. \citet{abdi2005distatis} propose the choice 
\begin{equation}\label{eq::frob_weights}
\mathbf{w}_\mathrm{F}=\frac{1}{\mathbf{1}^{\mathrm{T}}\mathbf{q}_1}\,\mathbf{q}_1.
\end{equation}
\noindent
In practice, a principal component analysis is considered on $\mathbf{R}$ and the first eigenvector is used for implementing $\mathbf{w}$. 
This choice reflects the idea that layers with larger projections on $\mathbf{q}_1$ are more similar to the other layers than the layers with smaller projections. Therefore, the elements of this eigenvector should provide suitable weights for $\mathbf{S}_{+\mathrm{F}}$, which is a linear combination of $\mathbf{S}_1,\ldots,\mathbf{S}_m$.

In the cases of $\mathbf{S}_{+\mathrm{R}}$ and $\mathbf{S}_{+\mathrm{W}}$, it is not obvious if the previous choice of $\mathbf{w}$ is suitable, since these averages are not linear functions of $\mathbf{S}_1,\ldots,\mathbf{S}_m$. Alternatively, \citet{baccini2022average} suggest the choice
\begin{equation}\label{eq::riem_weights}
    \mathbf{w}_\mathrm{R}=\mathbf{w}_\mathrm{W}=\frac{1}{\mathbf{1}^{\mathrm{T}}(\mathbf{R}-\mathbf{I}_m)\mathbf{1}}\,(\mathbf{R}-\mathbf{I}_m)\mathbf{1}.
\end{equation}
\noindent
The rationale behind the weights in Equations \ref{eq::frob_weights} and \ref{eq::riem_weights} is that the more a similarity matrix is correlated with the other layers, the more it is representative of the whole set of similarity matrices -- and hence it should receive a larger weight with respect to the others. As a result, groups of layers that are assigned similar weights collect similarity matrices that are likely to be structurally similar. 

We conclude by remarking that the matrices $\mathbf{S}_{+\mathrm{F}}$, $\mathbf{S}_{+\mathrm{R}}$ and $\mathbf{S}_{+\mathrm{W}}$ represent the adjacency matrices of the desired monoplex in the practical implementation of the method. We also remark that SMA could be generally adopted with similarity matrices defined in $\mathcal{P}_n$, since the above results also hold in this space \citep{baccini2022average}. However, even if the method could be adopted for similarity matrices belonging to different spaces, the described properties of SMA are not necessarily guaranteed.

\section{Application to the analysis of glacier data}\label{sec::3}

In cryoconite holes, bacteria form complex ecological networks \citep{gokul2016taxon} whereby different populations interact, sometimes using peculiar metabolic ways \citep{franzetti2016light, pittino2023a}. However, it is still unknown whether the same ecological relationships between the bacterial populations of cryoconite holes occur worldwide, or whether a biogeography of glacier bacterial communities exist \citep{darcy2011global, franzetti2013bacterial, pittino2023b}. 

In the following, we performed the similarity network aggregation analysis described in Section \ref{sec::methods} on a set of ecological networks constructed from the relative abundances of the bacterial populations sampled on nine glaciers. 

\subsection{Dataset description and processing}

The dataset consists of the relative abundances obtained from 16S rDNA metatranscriptomics of the bacterial communities sampled between 2013 and 2018 on 4 glaciers in the Alps (Forni, Cedec, and East Zebrù in the Ortles-Cevedale group in Italy, Morteratsch in the Bernina group in Switzerland) and five glaciers in the Andes (Iver in the Plomo Group, El Morado in the Maipo Valley, Exploradores in the San Valentin group in Chile, Perito Moreno in the Southern Patagonian Ice Field in Argentina). Details on glaciers, sampling methods, laboratory and bioinformatics analyses are provided by \citet{pittino2018bacterial, pittino2021diel}. Briefly, 16S rDNA was extracted from cryoconite collected in 140 holes (11-23 holes per glacier), Illumina sequenced and clustered in Amplicon Sequences Variants (ASVs) with DADA2 \citep{callahan2016dada2}. ASVs, which can roughly be considered to correspond to bacterial strains, were then taxonomically classified using the SILVA classifier \citep{SILVA2012}.
Overall, the dataset consists of $135$ ASVs, where the number of cryoconite holes considered varies between glaciers. Table \ref{tab::numberofsitesperglac} reports the exact number of sites (cryoconite holes) considered for each glacier. 

\begin{table}[!ht]
    \centering
        \caption{Number of sites where the measurement of relative abundance of the considered ASVs was performed.}
    \begin{tabular}{c c}
    \hline
       Glacier  & \# Cryoconite holes \\
    \hline
       Cedec  & 11 \\
       Exploradores & 15 \\
       Forni & 20 \\
       Iver & 15 \\
       Iver Est & 15 \\
       El Morado & 15 \\
       Morteratsch & 23 \\
       Perito Moreno & 15 \\
       Zebr\'{u} & 11 \\
       \hline
    \end{tabular}
    \label{tab::numberofsitesperglac}
\end{table}

For the present analysis, the ASVs that are not present on any glacier are filtered out. This excludes $9$ of the $135$ ASVs available. In view of computing similarities between pairs of ASVs for each glacier, the ASVs that are totally absent in at least one glacier are filtered out. Indeed, the complete absence of an ASV is represented as a null vector of dimension equal to the number of different sites of the glacier. This constitutes an issue when computing the similarity between such ASV and each of the remaining ones, since a null vector has similarity equal to $0$ with all the other vectors, and produces a similarity matrix which is not completely positive. As a consequence, the integration methods described in Section \ref{sec::methods} could not be applied. By excluding the ASVs that are absent from at least one glacier, the number of ASVs drastically reduces to $16$, represented by $1,626,370$ sequences. The taxonomy of the considered ASVs is reported in Table \ref{tab::taxonomy}. We highlight that this filtering is acceptable if we focus the attention on the most common ASVs that can be found on all the analysed glaciers. A more detailed analysis of the excluded ASVs goes beyond the scope of the present work, and is left for future investigations.

Overall, the filtering process yields $9$ relative abundance matrices, one for each glacier, where one dimension is fixed and equal to $16$ (the number of ASVs) and the other varies according to the number of  cryoconite holes sampled on each glacier (see Table \ref{tab::numberofsitesperglac}).

\begin{sidewaystable}
    \centering
    \caption{Taxonomy of the 16 ASVs included in the analysis. The ASV number is assigned in the data collection phase and it is unique for each sample.\\}
    \resizebox{\textwidth}{!}{
    \begin{tabular}{c c c c c c c}
    \hline
        ASV & Domain & Phylum & Class & Order & Family & Genus \\ 
    \hline
        ASV\_166579 & Bacteria & Bacteroidetes & Sphingobacteriia & Sphingobacteriales & Chitinophagaceae & \textit{Ferruginibacter} \\ 
        ASV\_179530 & Bacteria & Proteobacteria & Gammaproteobacteria & Pseudomonadales & Pseudomonadaceae & \textit{Rhizobacter} \\ 
        ASV\_184088 & Bacteria & Proteobacteria & Betaproteobacteria & Burkholderiales & Comamonadaceae & \textit{Acidovorax} \\ 
        ASV\_184128 & Bacteria & Proteobacteria & Betaproteobacteria & Burkholderiales & Comamonadaceae & \textit{Variovorax} \\ 
        ASV\_184156 & Bacteria & Proteobacteria & Betaproteobacteria & Burkholderiales & Comamonadaceae & \textit{Unclassified\_Comamonadaceae} \\ 
        ASV\_185564 & Bacteria & Proteobacteria & Betaproteobacteria & Burkholderiales & Comamonadaceae & \textit{Polaromonas} \\
        ASV\_185576 & Bacteria & Proteobacteria & Betaproteobacteria & Burkholderiales & Comamonadaceae & \textit{Polaromonas} \\ 
        ASV\_185593 & Bacteria & Proteobacteria & Betaproteobacteria & Burkholderiales & Comamonadaceae & \textit{Polaromonas} \\ 
        ASV\_190905 & Bacteria & Proteobacteria & Betaproteobacteria & Burkholderiales & Oxalobacteraceae & \textit{Unclassified\_Oxalobacteraceae} \\ 
        ASV\_218814 & Bacteria & Proteobacteria & Alphaproteobacteria & Rhizobiales & Bradyrhizobiaceae & \textit{Bradyrhizobium} \\ 
        ASV\_310446 & Bacteria & Bacteroidetes & Cytophagia & Cytophagales & Cytophagaceae & \textit{Hymenobacter} \\ 
        ASV\_310875 & Bacteria & Bacteroidetes & Cytophagia & Cytophagales & Cytophagaceae & \textit{Hymenobacter} \\ 
        ASV\_439437 & Bacteria & Proteobacteria & Alphaproteobacteria & Rhizobiales & Phyllobacteriaceae & \textit{Mesorhizobium} \\ 
        ASV\_489786 & Bacteria & Actinobacteria & Actinobacteria & Actinomycetales & Microbacteriaceae & \textit{Cryobacterium} \\ 
        ASV\_88033 & Bacteria & Cyanobacteria/Chloroplast & Cyanobacteria & Unclassified\_Cyanobacteria & Unclassified\_Cyanobacteria & \textit{Unclassified\_Cyanobacteria} \\ 
        ASV\_92579 & Bacteria & Deinococcus-Thermus & Deinococci & Deinococcales & Deinococcaceae & \textit{Deinococcus} \\ 
        \hline
    \end{tabular}}
    \label{tab::taxonomy}
\end{sidewaystable}
\subsection{Integration of the multiplex similarity network of ASVs}\label{sec::integrationapp}

In order to assess the existence of differences between glaciers in terms of the relative abundance of ASVs, a multiplex similarity network of ASVs is computed. In particular, the aim is to select the ASVs that are present in every glacier, and then compute the similarities between pairs of ASVs based on their relative abundance within every glacier. The result is a multiplex similarity network where each layer represents a glacier, and the nodes are the considered ASVs. 
All the ASVs included in the analysis belong to orders and genera typical of cryoconite holes worldwide and, in particular, of those situated on mountain glaciers \citep{fillinger2021spatial}. 
Formally, the $9$ matrices can be interpreted as a multilayer network formed by $9$ bipartite networks with the structure discussed at the beginning of Section \ref{sec::methods}. Indeed, each matrix is the adjacency matrix of a bipartite network, where the nodes in one set of the bipartition represent the ASVs, and the other set varies between layers and represents the set of sites. We remark that the set of nodes representing the ASVs is the same on each layer. Moreover, a node corresponding to an ASV is connected to all the nodes corresponding to sites where the ASV is present. These links are weighted with the relative abundance value of that ASV in the connected site. Subsequently, the similarity between pairs of ASVs is computed as in \eqref{eq::wandsim}, and a multiplex similarity network is obtained.

\begin{sidewaysfigure}
    \includegraphics[width=\textwidth]{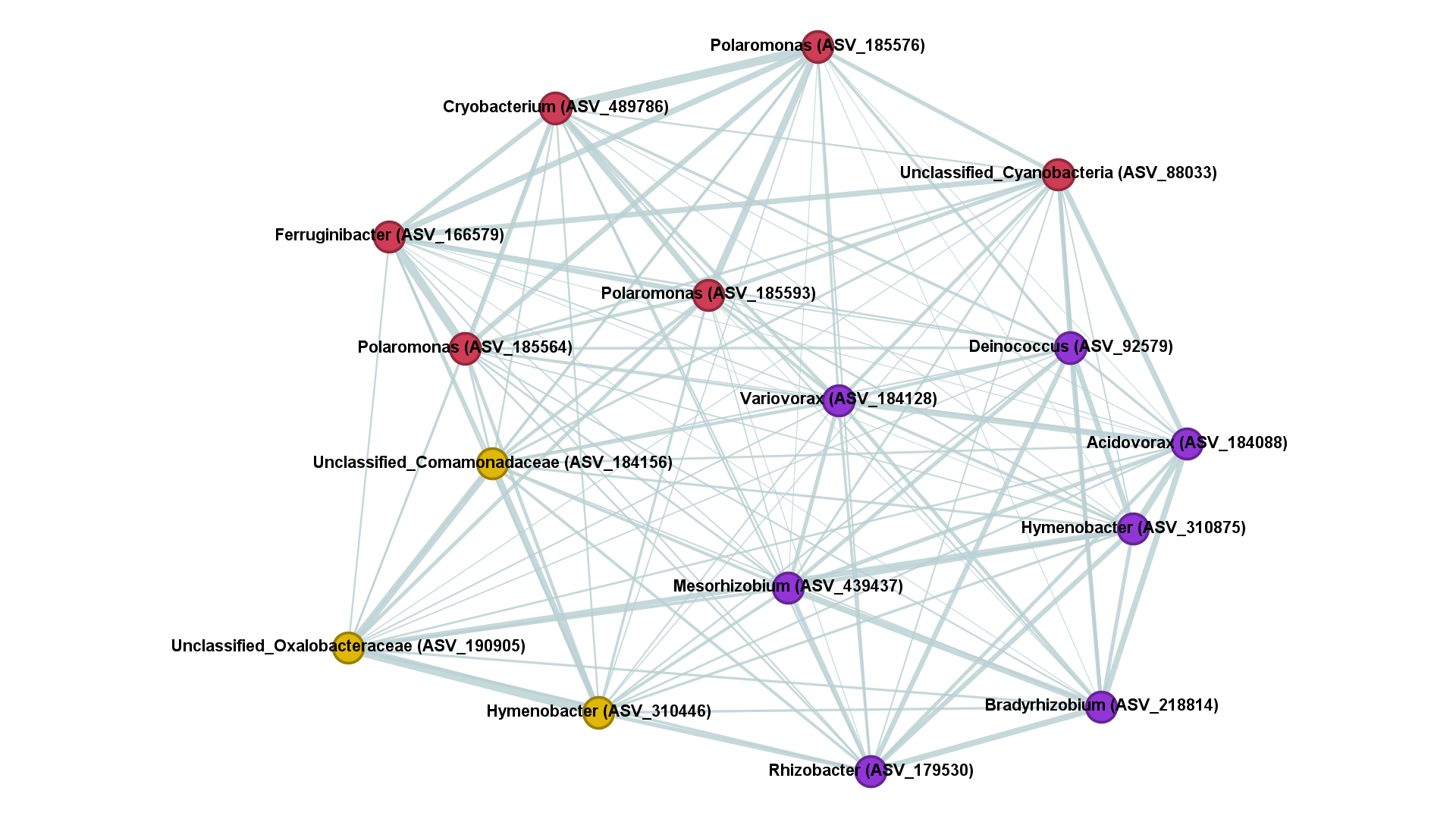}
    \caption{Monoplex obtained with SNF ($\mathbf{S}_{\text{SNF}}$). Different colours correspond to distinct communities individuated by the Louvain algorithm for modularity optimization. Node labels report the genus and the identification number of the corresponding ASVs.}
    \label{fig::SNFnet}
\end{sidewaysfigure}
\begin{sidewaysfigure}
    \includegraphics[width=\textwidth]{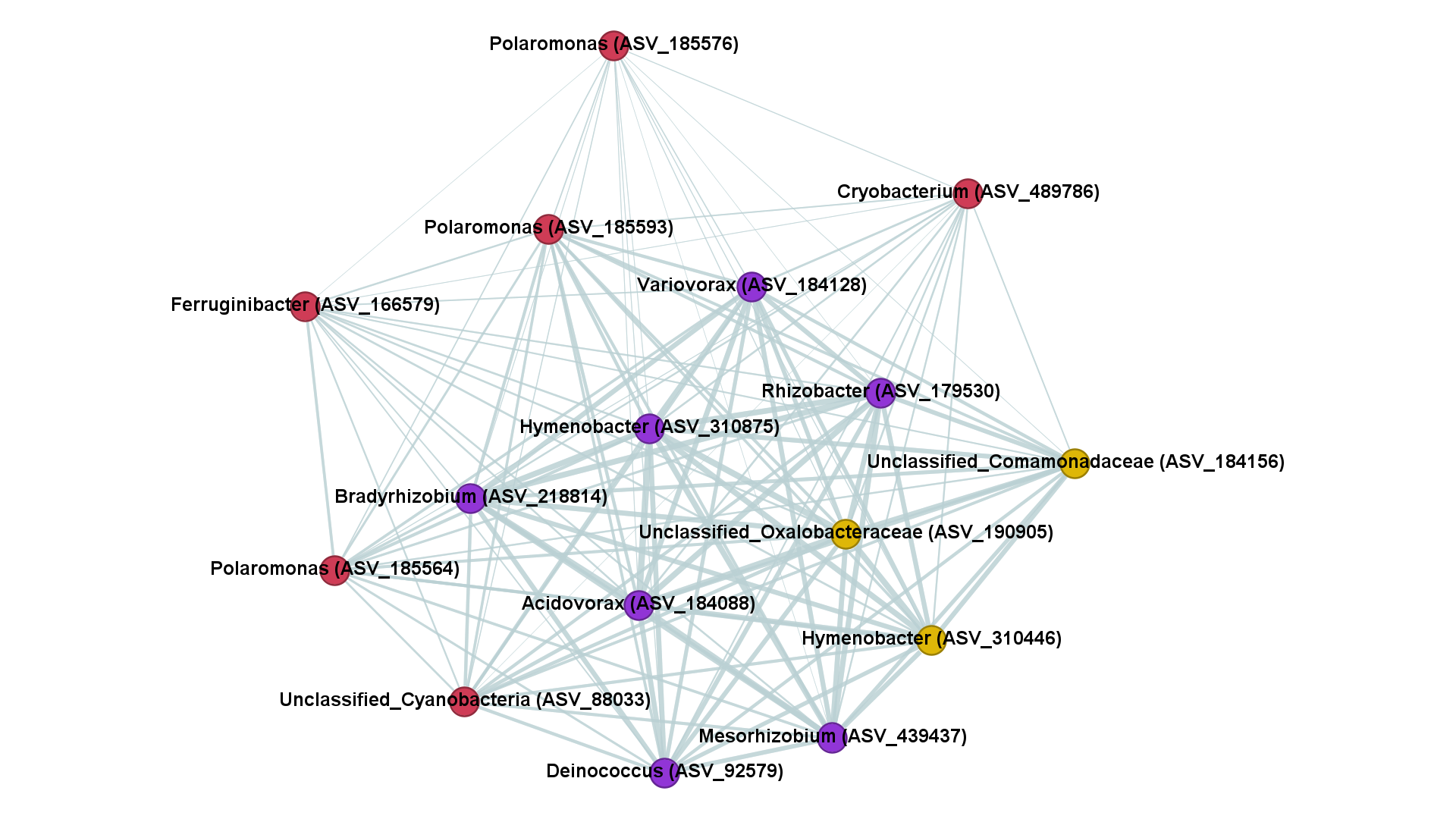}
    \caption{Monoplex obtained with SMA and the Frobenius metric ($\mathbf{S}_{\text{+F}}$). Different colours correspond to distinct communities individuated by the Louvain algorithm for modularity optimization. Node labels report the genus and the identification number of the corresponding ASVs.}
    \label{fig::frobnet}
\end{sidewaysfigure}
\begin{sidewaysfigure}
    \includegraphics[width=\textwidth]{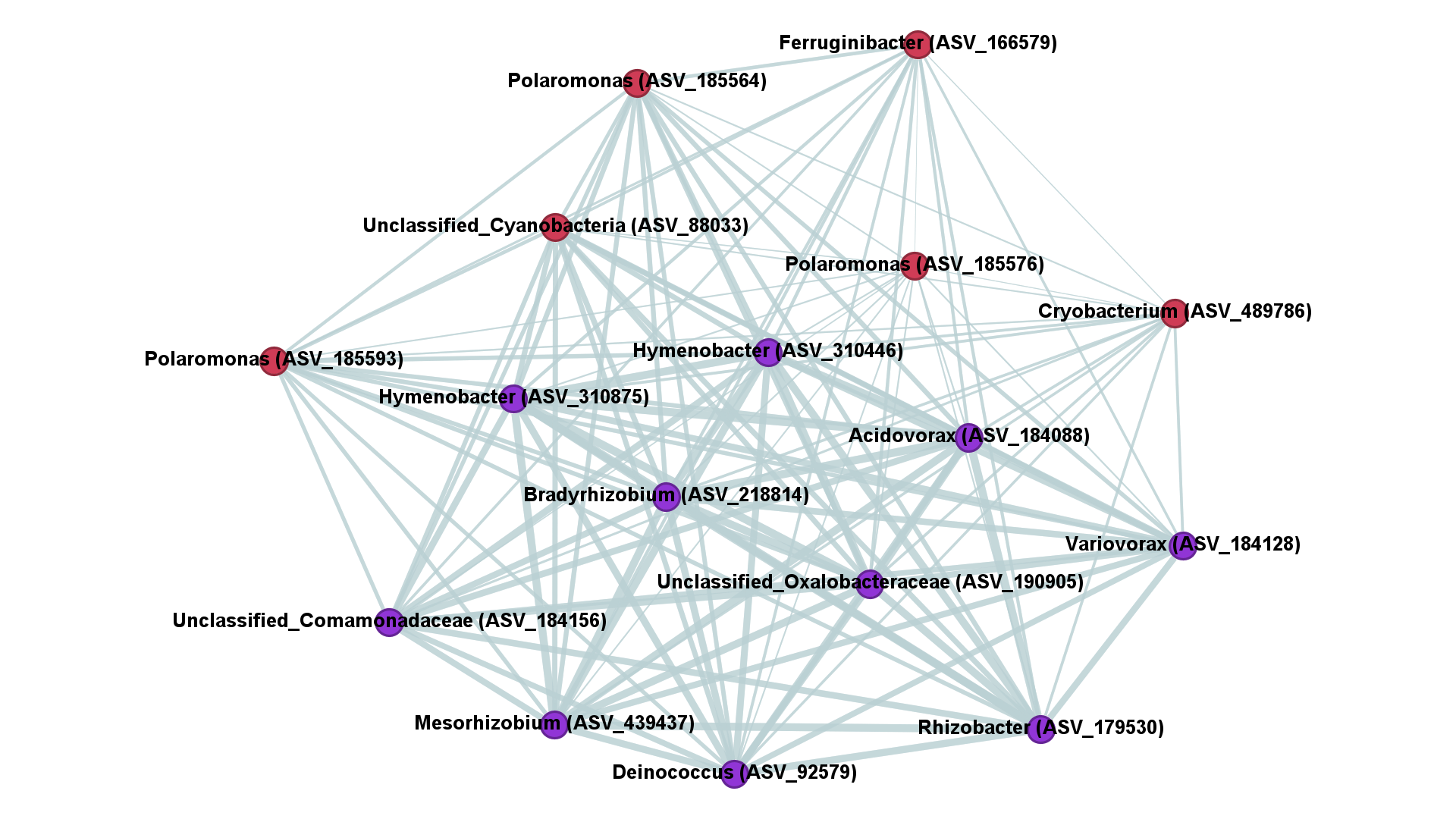}
    \caption{Monoplex obtained with SMA and the Riemannian metric ($\mathbf{S}_{\text{+R}}$). Different colours correspond to distinct communities individuated by the Louvain algorithm for modularity optimization. Node labels report the genus and the identification number of the corresponding ASVs.}
    \label{fig::riemnet}
\end{sidewaysfigure}
\begin{sidewaysfigure}
    \includegraphics[width=\textwidth]{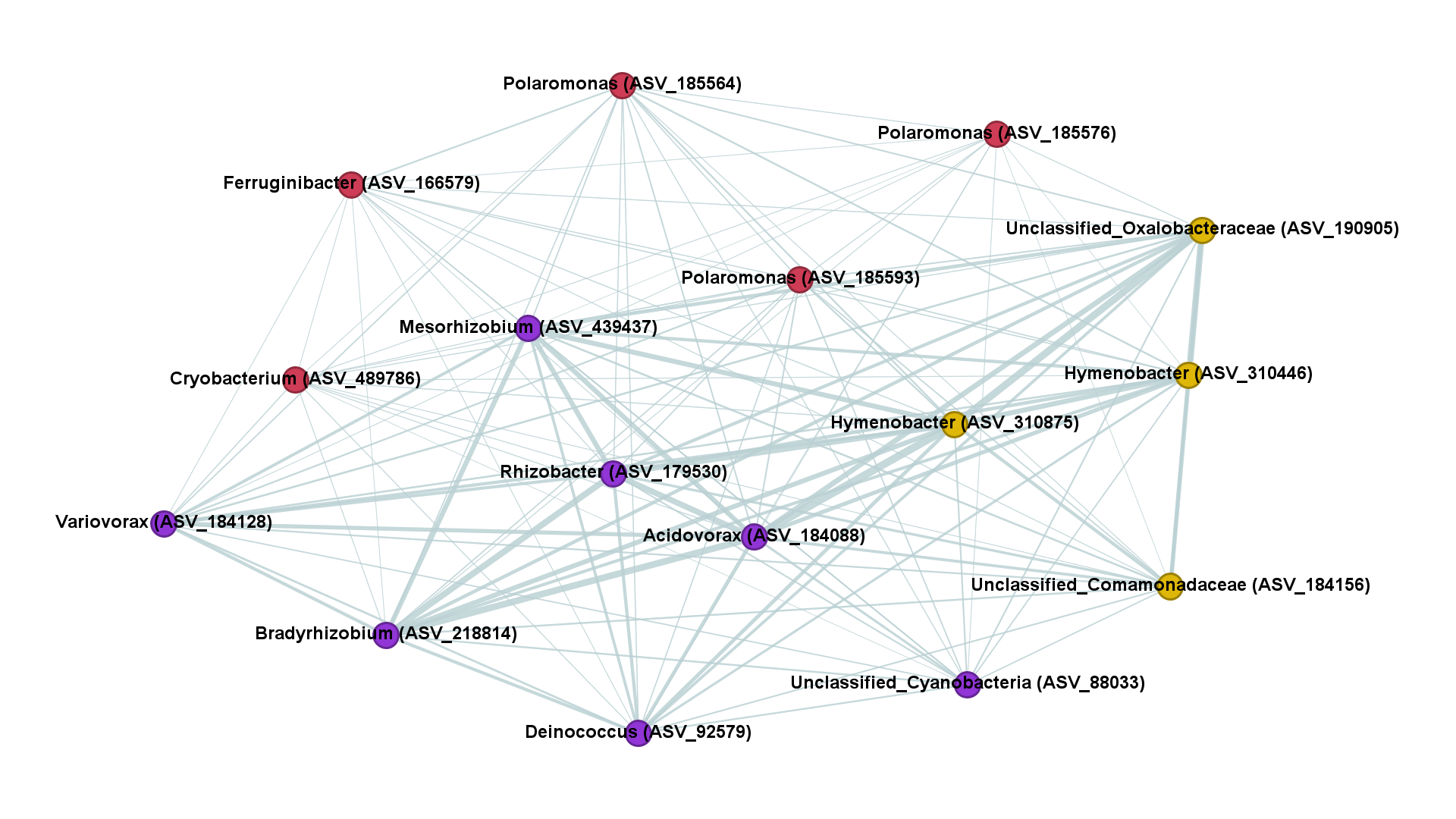}
    \caption{Monoplex obtained with SMA and the Wasserstein metric ($\mathbf{S}_{\text{+W}}$). Different colours correspond to distinct communities individuated by the Louvain algorithm for modularity optimization. Node labels report the genus and the identification number of the corresponding ASVs.}
    \label{fig::wassnet}
\end{sidewaysfigure}

In order to analyse the overall organization of the considered ASVs, both SNF and SMA are applied to the multiplex similarity network of ASVs. To perform the integration we used the \proglang{R} implementation of SNF, available at \url{http://compbio.cs.toronto.edu/SNF/SNF/Software.html}, and the \proglang{Python} implementation of SMA available at \url{https://github.com/DedeBac/SimilarityMatrixAggregation}. Four different monoplex networks are obtained: one resulting from SNF, i.e.\ $\mathbf{S}_\mathrm{SNF}$, and one for each of the three different metric choices available in SMA and discussed in Section \ref{subsec::sma}, i.e\ $\mathbf{S}_{+\mathrm F}$, $\mathbf{S}_{+\mathrm R}$ and $\mathbf{S}_{+\mathrm W}$. For each resulting network, the Louvain algorithm for modularity optimization \citep{blondel2008modularity} is then applied to individuate strongly connected groups of nodes. The networks are depicted in Figures \ref{fig::SNFnet}-\ref{fig::wassnet}. In the figures, the nodes are coloured according to the cluster they have been assigned to. The cluster and visual analysis was performed using the Gephi software \citep{bastian2009gephi}.

All the ASVs included in the analysis belong to orders and genera typical of cryoconite holes worldwide, particularly on mountain glaciers \citep{pittino2023b}. Except for $\mathbf{S}_{+\mathrm R}$, for which only two clusters are individuated, the Louvain algorithm partitions the networks into three clusters. In all the networks, the obtained modules consistently identified a group (red nodes) of ASVs composed by the genera \textit{Polaromonas} (3 ASVs), \textit{Cryobacterium}, \textit{Ferruginibacter} and \textit{Cyanobacteria} (in $\mathbf{S}_{+\mathrm W}$ \textit{Cyanobacteria} are instead assigned to the purple cluster). Cyanobacteria usually dominate cryoconite hole bacterial communities, where they are typically primary producers and ecosystem engineers \citep{rozwalak2022cryoconite}. The other taxa belonging to the red cluster show a broad variety of metabolisms, which allow them to survive on a broad variety of energy sources in the harsh supraglacial environment \citep{franzetti2013bacterial}. Overall, the red cluster is composed of bacterial strains that are abundant and typical of cryoconite holes worldwide \citep{boetius2015microbial, fillinger2021spatial, franzetti2013bacterial}.  

In the network obtained with SNF, in $\mathbf{S}_{+\mathrm F}$ and in $\mathbf{S}_{+\mathrm W}$ (Figures \ref{fig::SNFnet}, \ref{fig::frobnet} and \ref{fig::wassnet}) the Louvain algorithm identifies a further module (yellow nodes) composed of one ASV classified as \textit{Himenobacter}, and two ASVs classified only at the family level as \textit{Comamonadaceae} and \textit{Oxalobacteraceae} (see Table \ref{tab::taxonomy}). In $\mathbf{S}_{+\mathrm W}$ a second Himenobacter ASV is also added to the yellow cluster. However, the lack of classification at the genus level and the inconsistency in the classification of the Himenobacter ASVs prevent from further interpretations of the ecological role of the yellow cluster in the cryoconite hole communities. 

\newpage
\subsection{Results and discussion}\label{subsec::results}

The first issue to be investigated is the coherence between the various monoplex networks obtained with different techniques. To this aim, a measure of correlation between the similarity matrices (i.e.\ the similarity networks) is adopted. An appropriate correlation measure is the so-called ``generalized distance correlation" introduced by \citet{szekely2007measuring}. In the present setting, the generalized distance correlation can be interpreted as an extension to similarity matrices of the Pearson correlation coefficient, since it assumes values in the interval $[0,1]$, with values close to $0$ indicating no or very weak association, and values close to $1$ suggesting a stronger association. Table \ref{tab::dcor_monoplex} displays the values of generalized distance correlation between pairs of monoplex networks obtained with different methods. It is apparent that all the networks are strongly associated, thus indicating a significant structural similarity between them. Therefore, this argument suggests a coherence between the different methods.  

\begin{table}[!ht]
    \centering
        \caption{Values of the generalized distance correlation between monoplex networks obtained with different methods. 
        \newline}
    \begin{tabular}{c c c c c}
    \hline
         & $\mathbf{S}_\mathrm{SNF}$ & $\mathbf{S}_{+\mathrm F}$ & $\mathbf{S}_{+\mathrm R}$ & $\mathbf{S}_{+\mathrm W}$\\
     \hline
        $\mathbf{S}_\mathrm{SNF}$ & $1$ & $0.82659$ & $0.78627$ & $0.99790$ \\
         $\mathbf{S}_{+\mathrm F}$ & $ - $ & $1$   & $0.99000$ & $0.83454$ \\
         $\mathbf{S}_{+\mathrm R}$ & $ - $ & $ - $  & $1$ & $0.79449$ \\
         $\mathbf{S}_{+\mathrm W}$ & $ - $ & $ - $ & $ - $ & $ 1 $ \\
     \hline
    \end{tabular}
    \label{tab::dcor_monoplex}
\end{table}

The second issue of interest is the relative impact of the single layers (i.e.\ the glaciers) in determining the final monoplex. As highlighted in Section \ref{subsec::sma}, unlike SNF, SMA automatically assigns weights to the contribution of each layer to the final network by using the RV-coefficients matrix. The resulting weights may be interpreted as the relative importance of the single layers for the determination of the final monoplex. Moreover, networks (in this case, glaciers) that are assigned similar weights are likely to have a similar structure. 
Table \ref{tab::weights} reports the weights assigned to each layer (i.e.\ the glacier) in the SMA procedure, in case of the adoption of the different choices given by \eqref{eq::frob_weights} and \eqref{eq::riem_weights}.

\begin{table}[!ht]
    \centering
    \caption{Weights assigned to the similarity networks of the 9 different layers. \newline }
    \medskip
    \begin{tabular}{c c c}
    \hline
        Glacier & $\mathbf{w}_\mathrm{F}$ & $\mathbf{w}_\mathrm{R}$\\
    \hline
        Cedec & $0.11333$ & $0.11355$\\ 
        Exploradores & $0.10664$ & $0.10616$\\
        Forni & $0.11026$ & $0.11009$\\
        Iver & $0.11003$ & $0.10990$\\
        Iver Est & $0.11418$ & $0.11459$\\
        El Morado & $0.11424$ & $0.11463$\\
        Morteratsch & $0.11043$ & $0.11037$\\
        Perito Moreno & $0.10787$ & $0.10749$\\
        Zebr\'{u} & $0.11301$ & $0.11321$\\
    \hline
    \end{tabular}
    \label{tab::weights}
\end{table}

From Table \ref{tab::weights} it can be observed that Exploradores and Perito Moreno are assigned very similar weights, both slightly lower than those assigned to the other glaciers. Interestingly, Exploradores and Perito Moreno are large glaciers ($> 85$ km$^2$) reaching low altitudes (the sampling areas were $\le$ 200 m a.s.l.), while the other glaciers (both in the Andes and the Alps) are medium to small ($\le$ 15 km$^2$) mountain glaciers at $\ge$ 2500 m a.s.l. Thus, Exploradores and Perito Moreno can show very different ecological conditions and communities with different structures with respect to the other glaciers, which are reflected by their similar weights. 

As previously observed, SNF does not assign specific weights to the contribution of the single layers. However, it is possible, a posteriori, to adopt suitable network correlation measures to evaluate the structural similarity between the final monoplex and the single layers. An appropriate choice of such measure is the generalized distance correlation previously discussed. Table \ref{tab::dcorSNF} reports the values of distance correlation between the monoplex obtained with SNF and the single similarity networks corresponding to different glaciers. Coherently with the weights assigned by the SMA algorithm, Exploradores and Perito Moreno show a weaker association with the final monoplex, when compared with the other glaciers. 

\begin{table}[!ht]
    \centering
        \caption{Values of the generalized distance correlation between the monoplex obtained with SNF and the single layers.}
        \medskip
    \begin{tabular}{c c}
    \hline
        Glacier & \\
    \hline
        Cedec & $0.74192$ \\ 
        Exploradores & $0.67823$ \\
        Forni & $0.84335$ \\
        Iver & $0.74968$\\
        Iver Est & $0.71703$\\
        El Morado & $0.82913$\\
        Morteratsch & $0.87491$\\
        Perito Moreno & $0.68240$\\
        Zebr\'{u} & $0.70132$\\
    \hline
    \end{tabular}
    \label{tab::dcorSNF}
\end{table}

\section{Conclusion}\label{sec::conclusion}

The present paper discusses the use of two network integration methods, namely Similarity Network Fusion and Similarity Matrix Average, to combine multi-source information coming from ecological networks into a single network. The two methods are valid alternatives to integrate a multiplex similarity network, and are shown to give coherent results. A major difference between SNF and SMA is that the second method includes an automatic way of assigning weights to the contribution of the layers based on the structure of the multiplex. In contrast, when SNF is used, the impact of the layers in the final monoplex can only be evaluated a posteriori by computing, for instance, the generalized distance correlation. 
The effectiveness of the two methods in the analysis of ecological networks is tested on a real-world dataset recording the relative abundance of microbial species in the ecosystems of nine glaciers. Despite being based on a very small subset of all the ASVs observed on the considered glaciers (19 out of 126), the investigation was able to identify an ecological network between cosmopolitan glacier bacteria strains, which may form the “basic” ecological network of glacier ecosystems. Further investigation on a larger set of glaciers will be certainly necessary to assess the generality of this ecological network. Nevertheless, the present analysis can provide glacier ecologists with a background hypothesis for future investigations. Moreover, the weights assigned to the similarity networks and the generalized distance correlations between them consistently suggest that the two large Patagonian glaciers (Exploradores and Perito Moreno) host ecological networks different from those of the other smaller mountain glaciers. This pattern supports the hypothesis that the general features of a glacier, as its altitude, size, and whether or not its tongue crosses the local tree line, can affect their ecological networks \citep{pittino2021diel}.  
Overall, this work opens to the use and development of similarity network integration techniques to combine different characteristics describing biological entities with the aim of discovering organization patterns in ecological communities. 

\section*{Declarations}
\begin{itemize}
\item Conflict of interest/Competing interests: The authors have no competing interests to declare that are relevant to the content of this article.
\item Availability of data and materials: the data used in this study are available from the corresponding author upon reasonable request. 
\item Authors' contributions: Federica Baccini and Lucio Barabesi contributed to the analysis and developements of the methodology. Roberto Ambrosini collected the data. Federica Baccini performed the data analysis and the computation of the matrix integration methods. Roberto Ambrosini provided the data and interpreted them from an ecological perspective. All the authors contributed to the writing of the manuscript.
\end{itemize}

\section*{Acknowledgements}
The authors are grateful to Pablo Alviz-Gazitua, Roberto Sergio Azzoni, Biagio Di Mauro, Guglielmina Diolaiuti, and Michael Seeger for their help during field work. Andrea Franzetti and Francesca Pittino kindly provided sequencing data. We thank the administration of Parco Nazionale dello Stelvio for the permission of collecting samples on Cedec, Forni and Zebr\'u glaciers through their collaboration agreement with the University of Milan; the CONAF (Corporación Nacional Forestal of Chile) for the permission and support in collecting supraglacial samples on the Exploradores Glacier, and the Administración de Parques Nacionales of Argentina (APN) for the permission to collect samples on the Perito Moreno Glacier with project N. 110-CPA-2016. La Venta team provided logistic support during the expeditions.

\bibliographystyle{plainnat}
\bibliography{references}  

\begin{thebibliography}{48}
\providecommand{\natexlab}[1]{#1}
\providecommand{\url}[1]{\texttt{#1}}
\expandafter\ifx\csname urlstyle\endcsname\relax
  \providecommand{\doi}[1]{doi: #1}\else
  \providecommand{\doi}{doi: \begingroup \urlstyle{rm}\Url}\fi

\bibitem[Abdi et~al.(2005)Abdi, O'Toole, Valentin, and Edelman]{abdi2005distatis}
Herv{\'e} Abdi, Alice~J O'Toole, Dominique Valentin, and Betty Edelman.
\newblock {DISTATIS}: The analysis of multiple distance matrices.
\newblock In \emph{Proceedings of the IEEE Computer Society: International Conference on Computer Vision and Pattern Recognition, San Diego, CA}, pages 42--47, 2005.

\bibitem[{\'A}lvarez-Esteban et~al.(2016){\'A}lvarez-Esteban, Del~Barrio, Cuesta-Albertos, and Matr{\'a}n]{alvarez2016fixed}
Pedro~C {\'A}lvarez-Esteban, E~Del~Barrio, JA~Cuesta-Albertos, and C~Matr{\'a}n.
\newblock A fixed-point approach to barycenters in wasserstein space.
\newblock \emph{Journal of Mathematical Analysis and Applications}, 441\penalty0 (2):\penalty0 744--762, 2016.

\bibitem[Ambrosini et~al.(2017)Ambrosini, Musitelli, Navarra, Tagliaferri, Gandolfi, Bestetti, Mayer, Minora, Azzoni, Diolaiuti, et~al.]{ambrosini2017diversity}
Roberto Ambrosini, Federica Musitelli, Federico Navarra, Ilario Tagliaferri, Isabella Gandolfi, Giuseppina Bestetti, Christoph Mayer, Umberto Minora, Roberto~Sergio Azzoni, Guglielmina Diolaiuti, et~al.
\newblock Diversity and assembling processes of bacterial communities in cryoconite holes of a karakoram glacier.
\newblock \emph{Microbial ecology}, 73:\penalty0 827--837, 2017.

\bibitem[Anesio and Laybourn-Parry(2012)]{anesio2012glaciers}
Alexandre~M Anesio and Johanna Laybourn-Parry.
\newblock Glaciers and ice sheets as a biome.
\newblock \emph{Trends in ecology \& evolution}, 27\penalty0 (4):\penalty0 219--225, 2012.

\bibitem[Bac{\'a}k(2014)]{bacak2014computing}
Miroslav Bac{\'a}k.
\newblock Computing medians and means in {H}adamard spaces.
\newblock \emph{SIAM Journal on Optimization}, 24\penalty0 (3):\penalty0 1542--1566, 2014.

\bibitem[Baccini et~al.(2022{\natexlab{a}})Baccini, Barabesi, Baccini, Khelfaoui, and Gingras]{baccini2022similarity}
Federica Baccini, Lucio Barabesi, Alberto Baccini, Mahdi Khelfaoui, and Yves Gingras.
\newblock Similarity network fusion for scholarly journals.
\newblock \emph{Journal of Informetrics}, 16\penalty0 (1):\penalty0 101226, 2022{\natexlab{a}}.

\bibitem[Baccini et~al.(2022{\natexlab{b}})Baccini, Bianchini, and Geraci]{baccini2022graph}
Federica Baccini, Monica Bianchini, and Filippo Geraci.
\newblock Graph-based integration of histone modification profiles.
\newblock \emph{Mathematics}, 10\penalty0 (11):\penalty0 1842, 2022{\natexlab{b}}.

\bibitem[Baccini et~al.(2023)Baccini, Barabesi, and Petrovich]{baccini2022average}
Federica Baccini, Lucio Barabesi, and Eugenio Petrovich.
\newblock Similarity matrix average for aggregating multiplex networks.
\newblock \emph{Journal of Physics: Complexity}, 2023.
\newblock URL \url{http://iopscience.iop.org/article/10.1088/2632-072X/acda09}.

\bibitem[Barab{\'a}si(2013)]{barabasi2013network}
Albert-L{\'a}szl{\'o} Barab{\'a}si.
\newblock Network science.
\newblock \emph{Philosophical Transactions of the Royal Society A: Mathematical, Physical and Engineering Sciences}, 371\penalty0 (1987):\penalty0 20120375, 2013.

\bibitem[Bastian et~al.(2009)Bastian, Heymann, and Jacomy]{bastian2009gephi}
Mathieu Bastian, Sebastien Heymann, and Mathieu Jacomy.
\newblock Gephi: an open source software for exploring and manipulating networks.
\newblock In \emph{Proceedings of the international AAAI conference on web and social media}, volume~3, pages 361--362, 2009.

\bibitem[Berman and Shaked-Monderer(2003)]{berman2003completely}
Abraham Berman and Naomi Shaked-Monderer.
\newblock \emph{Completely Positive Matrices}.
\newblock World Scientific, Singapore, 2003.

\bibitem[Bhatia(2009)]{bhatia2009positive}
Rajendra Bhatia.
\newblock \emph{Positive Definite Matrices}.
\newblock Princeton University Press, Princeton, 2009.

\bibitem[Bhatia and Congedo(2019)]{bhatia2019procrustes}
Rajendra Bhatia and Marco Congedo.
\newblock Procrustes problems in {R}iemannian manifolds of positive definite matrices.
\newblock \emph{Linear Algebra and its Applications}, 563:\penalty0 440--445, 2019.

\bibitem[Bhatia et~al.(2019)Bhatia, Jain, and Lim]{bhatia2019bures}
Rajendra Bhatia, Tanvi Jain, and Yongdo Lim.
\newblock On the {B}ures--{W}asserstein distance between positive definite matrices.
\newblock \emph{Expositiones Mathematicae}, 37\penalty0 (2):\penalty0 165--191, 2019.

\bibitem[Bianconi(2018)]{bianconi2018multilayer}
Ginestra Bianconi.
\newblock \emph{Multilayer Networks}.
\newblock Oxford University Press, 2018.

\bibitem[Blondel et~al.(2008)Blondel, Guillaume, Lambiotte, and Lefebvre]{blondel2008modularity}
Vincent~D. Blondel, Jean-Loup Guillaume, Renaud Lambiotte, and Etienne Lefebvre.
\newblock Fast unfolding of communities in large networks.
\newblock \emph{Journal of Statistical Mechanics: Theory and Experiment}, 2008\penalty0 (10):\penalty0 P10008, 2008.
\newblock \doi{10.1088/1742-5468/2008/10/p10008}.

\bibitem[Boetius et~al.(2015)Boetius, Anesio, Deming, Mikucki, and Rapp]{boetius2015microbial}
Antje Boetius, Alexandre~M Anesio, Jody~W Deming, Jill~A Mikucki, and Josephine~Z Rapp.
\newblock Microbial ecology of the cryosphere: sea ice and glacial habitats.
\newblock \emph{Nature Reviews Microbiology}, 13\penalty0 (11):\penalty0 677--690, 2015.

\bibitem[Callahan et~al.(2016)Callahan, McMurdie, Rosen, Han, Johnson, and Holmes]{callahan2016dada2}
Benjamin~J Callahan, Paul~J McMurdie, Michael~J Rosen, Andrew~W Han, Amy Jo~A Johnson, and Susan~P Holmes.
\newblock Dada2: High-resolution sample inference from illumina amplicon data.
\newblock \emph{Nature methods}, 13\penalty0 (7):\penalty0 581--583, 2016.

\bibitem[Cook et~al.(2016)Cook, Edwards, Takeuchi, and Irvine-Fynn]{cook2016cryoconite}
Joseph Cook, Arwyn Edwards, Nozomu Takeuchi, and Tristram Irvine-Fynn.
\newblock Cryoconite: the dark biological secret of the cryosphere.
\newblock \emph{Progress in Physical Geography}, 40\penalty0 (1):\penalty0 66--111, 2016.

\bibitem[Darcy et~al.(2011)Darcy, Lynch, King, Robeson, and Schmidt]{darcy2011global}
John~L Darcy, Ryan~C Lynch, Andrew~J King, Michael~S Robeson, and Steven~K Schmidt.
\newblock Global distribution of polaromonas phylotypes-evidence for a highly successful dispersal capacity.
\newblock \emph{PLoS One}, 6\penalty0 (8):\penalty0 e23742, 2011.

\bibitem[De~Domenico(2022)]{de2022multilayer}
Manlio De~Domenico.
\newblock \emph{Multilayer Networks}.
\newblock Springer, 2022.

\bibitem[deBello et~al.(2017)deBello, Botta-Dukát, Lepš, and Fibich]{debello2021}
Francesco deBello, Zoltan Botta-Dukát, Jan Lepš, and Pavel Fibich.
\newblock Towards a more balanced combination of multiple traits when computing functional differences between species.
\newblock \emph{Methods in Ecology and Evolution}, 12:\penalty0 443--448, 2017.

\bibitem[Eck and Waltman(2009)]{eck2009normalize}
Nees Jan~van Eck and Ludo Waltman.
\newblock How to normalize cooccurrence data? an analysis of some well-known similarity measures.
\newblock \emph{Journal of the American society for information science and technology}, 60\penalty0 (8):\penalty0 1635--1651, 2009.

\bibitem[Estrada(2012)]{estrada2012structure}
Ernesto Estrada.
\newblock \emph{The Structure of Complex Networks}.
\newblock Oxford University Press, 2012.

\bibitem[Fillinger et~al.(2021)Fillinger, H{\"u}rkamp, Stumpp, Weber, Forster, Hausmann, Schultz, and Griebler]{fillinger2021spatial}
Lucas Fillinger, Kerstin H{\"u}rkamp, Christine Stumpp, Nina Weber, Dominik Forster, Bela Hausmann, Lotta Schultz, and Christian Griebler.
\newblock Spatial and annual variation in microbial abundance, community composition, and diversity associated with alpine surface snow.
\newblock \emph{Frontiers in Microbiology}, 12:\penalty0 781904, 2021.

\bibitem[Franzetti et~al.(2013)Franzetti, Tatangelo, Gandolfi, Bertolini, Bestetti, Diolaiuti, D'agata, Mihalcea, Smiraglia, and Ambrosini]{franzetti2013bacterial}
Andrea Franzetti, Valeria Tatangelo, Isabella Gandolfi, Valentina Bertolini, Giuseppina Bestetti, Guglielmina Diolaiuti, Carlo D'agata, Claudia Mihalcea, Claudio Smiraglia, and Roberto Ambrosini.
\newblock Bacterial community structure on two alpine debris-covered glaciers and biogeography of polaromonas phylotypes.
\newblock \emph{The ISME journal}, 7\penalty0 (8):\penalty0 1483--1492, 2013.

\bibitem[Franzetti et~al.(2016)Franzetti, Tagliaferri, Gandolfi, Bestetti, Minora, Mayer, Azzoni, Diolaiuti, Smiraglia, and Ambrosini]{franzetti2016light}
Andrea Franzetti, Ilario Tagliaferri, Isabella Gandolfi, Giuseppina Bestetti, Umberto Minora, Christoph Mayer, Roberto~S Azzoni, Guglielmina Diolaiuti, Claudio Smiraglia, and Roberto Ambrosini.
\newblock Light-dependent microbial metabolisms drive carbon fluxes on glacier surfaces.
\newblock \emph{The ISME journal}, 10\penalty0 (12):\penalty0 2984--2988, 2016.

\bibitem[Gokul et~al.(2016)Gokul, Hodson, Saetnan, Irvine-Fynn, Westall, Detheridge, Takeuchi, Bussell, Mur, and Edwards]{gokul2016taxon}
Jarishma~K Gokul, Andrew~J Hodson, Eli~R Saetnan, Tristram~DL Irvine-Fynn, Philippa~J Westall, Andrew~P Detheridge, Nozomu Takeuchi, Jennifer Bussell, Luis~AJ Mur, and Arwyn Edwards.
\newblock Taxon interactions control the distributions of cryoconite bacteria colonizing a high arctic ice cap.
\newblock \emph{Molecular ecology}, 25\penalty0 (15):\penalty0 3752--3767, 2016.

\bibitem[Hastie et~al.(2009)Hastie, Tibshirani, Friedman, and Friedman]{hastie2009elements}
Trevor Hastie, Robert Tibshirani, Jerome~H Friedman, and Jerome~H Friedman.
\newblock \emph{The Elements of Statistical Learning}, volume~2.
\newblock Springer, 2009.

\bibitem[Hodson et~al.(2008)Hodson, Anesio, Tranter, Fountain, Osborn, Priscu, Laybourn-Parry, and Sattler]{hodson2008glacial}
Andy Hodson, Alexandre~M Anesio, Martyn Tranter, Andrew Fountain, Mark Osborn, John Priscu, Johanna Laybourn-Parry, and Birgit Sattler.
\newblock Glacial ecosystems.
\newblock \emph{Ecological monographs}, 78\penalty0 (1):\penalty0 41--67, 2008.

\bibitem[Hutchinson et~al.(2019)Hutchinson, Bramon~Mora, Pilosof, Barner, Kéfi, Thébault, Jordano, and Stouffer]{hutchinson2019}
Matthew~C. Hutchinson, Bernat Bramon~Mora, Shai Pilosof, Allison~K. Barner, Sonia Kéfi, Elisa Thébault, Pedro Jordano, and Daniel~B. Stouffer.
\newblock Seeing the forest for the trees: Putting multilayer networks to work for community ecology.
\newblock \emph{Functional Ecology}, 33\penalty0 (2):\penalty0 206--217, 2019.

\bibitem[Kivel{\"a} et~al.(2014)Kivel{\"a}, Arenas, Barthelemy, Gleeson, Moreno, and Porter]{kivela2014multilayer}
Mikko Kivel{\"a}, Alex Arenas, Marc Barthelemy, James~P Gleeson, Yamir Moreno, and Mason~A Porter.
\newblock Multilayer networks.
\newblock \emph{Journal of Complex Networks}, 2\penalty0 (3):\penalty0 203--271, 2014.

\bibitem[Lim and P{\'a}lfia(2014{\natexlab{a}})]{lim2014weighted}
Yongdo Lim and Mikl{\'o}s P{\'a}lfia.
\newblock Weighted inductive means.
\newblock \emph{Linear Algebra and its Applications}, 453:\penalty0 59--83, 2014{\natexlab{a}}.

\bibitem[Lim and P{\'a}lfia(2014{\natexlab{b}})]{lim2014weighteda}
Yongdo Lim and Mikl{\'o}s P{\'a}lfia.
\newblock Weighted deterministic walks for the least squares mean on {H}adamard spaces.
\newblock \emph{Bulletin of the London Mathematical Society}, 46\penalty0 (3):\penalty0 561--570, 2014{\natexlab{b}}.

\bibitem[Mittelbach and McGill(2019)]{mittelbach2019}
Gary~G Mittelbach and Brian~J McGill.
\newblock \emph{Community Ecology, 2nd ed.}
\newblock Oxford University Press, 2019.

\bibitem[Newman(2018)]{newman2018networks}
Mark Newman.
\newblock \emph{Networks}.
\newblock Oxford University Press, 2018.

\bibitem[Pilosof et~al.(2017)Pilosof, Porter, Pascual, and Kéfi]{pilosof2017}
Shai Pilosof, Mason~A. Porter, Mercedes Pascual, and Sonia Kéfi.
\newblock The multilayer nature of ecological networks.
\newblock \emph{Nature Ecology \& Evolution}, 1\penalty0 (4):\penalty0 0101, 2017.

\bibitem[Pittino et~al.(2018)Pittino, Rossi, Ambrosini, Azzoni, Gandolfi, Diolaiuti, Franzetti, et~al.]{pittino2018bacterial}
F~Pittino, M~Rossi, R~Ambrosini, R~Azzoni, I~Gandolfi, G~Diolaiuti, A~Franzetti, et~al.
\newblock Bacterial community dynamics in cryoconite holes on alpine mountain glaciers.
\newblock 2018.

\bibitem[Pittino et~al.(2021)Pittino, Zordan, Azzoni, Diolaiuti, Ambrosini, and Franzetti]{pittino2021diel}
Francesca Pittino, Simone Zordan, Roberto~S Azzoni, Guglielmina Diolaiuti, Roberto Ambrosini, and Andrea Franzetti.
\newblock Diel transcriptional pattern contributes to functional and taxonomic diversity in supraglacial microbial communities.
\newblock \emph{bioRxiv}, pages 2021--01, 2021.

\bibitem[Pittino et~al.(2023{\natexlab{a}})Pittino, Ambrosini, Seeger, Azzoni, Diolaiuti, Alviz~Gazitua, and Franzetti]{pittino2023b}
Francesca Pittino, Roberto Ambrosini, Michael Seeger, Roberto~Sergio Azzoni, Guglielmina Diolaiuti, P~Alviz~Gazitua, and Andrea Franzetti.
\newblock Geographical variability of bacterial communities of cryoconite holes of andean glaciers.
\newblock \emph{Scientific Reports}, 13\penalty0 (1):\penalty0 2633, 2023{\natexlab{a}}.

\bibitem[Pittino et~al.(2023{\natexlab{b}})Pittino, Zawierucha, Poniecka, Buda, Rosatelli, Zordan, Azzoni, Diolaiuti, Ambrosini, and Franzetti]{pittino2023a}
Francesca Pittino, Krzysztof Zawierucha, Ewa Poniecka, Jakub Buda, Asia Rosatelli, Simone Zordan, Roberto~S Azzoni, Guglielmina Diolaiuti, Roberto Ambrosini, and Andrea Franzetti.
\newblock Functional and taxonomic diversity of anaerobes in supraglacial microbial communities.
\newblock \emph{Microbiology Spectrum}, 11\penalty0 (2):\penalty0 e01004--22, 2023{\natexlab{b}}.

\bibitem[Quast et~al.(2012)Quast, Pruesse, Yilmaz, Gerken, Schweer, Yarza, Peplies, and Gl{\"o}ckner]{SILVA2012}
Christian Quast, Elmar Pruesse, Pelin Yilmaz, Jan Gerken, Timmy Schweer, Pablo Yarza, J{\"o}rg Peplies, and Frank~Oliver Gl{\"o}ckner.
\newblock The silva ribosomal rna gene database project: improved data processing and web-based tools.
\newblock \emph{Nucleic acids research}, 41\penalty0 (D1):\penalty0 D590--D596, 2012.

\bibitem[Robert and Escoufier(1976)]{robert1976unifying}
Paul Robert and Yves Escoufier.
\newblock A unifying tool for linear multivariate statistical methods: the \it{{RV}}\rm{-coefficient}.
\newblock \emph{Journal of the Royal Statistical Society: Series C (Applied Statistics)}, 25\penalty0 (3):\penalty0 257--265, 1976.

\bibitem[Rozwalak et~al.(2022)Rozwalak, Podkowa, Buda, Niedzielski, Kawecki, Ambrosini, Azzoni, Baccolo, Ceballos, Cook, et~al.]{rozwalak2022cryoconite}
Piotr Rozwalak, Pawe{\l} Podkowa, Jakub Buda, Przemys{\l}aw Niedzielski, Szymon Kawecki, Roberto Ambrosini, Roberto~S Azzoni, Giovanni Baccolo, Jorge~L Ceballos, Joseph Cook, et~al.
\newblock Cryoconite--from minerals and organic matter to bioengineered sediments on glacier's surfaces.
\newblock \emph{Science of The Total Environment}, 807:\penalty0 150874, 2022.

\bibitem[Shaked-Monderer and Berman(2021)]{shaked2021copositive}
Naomi Shaked-Monderer and Abraham Berman.
\newblock \emph{Copositive and Completely Positive Matrices}.
\newblock World Scientific, Singapore, 2021.

\bibitem[Székely et~al.(2007)Székely, Rizzo, and Bakirov]{szekely2007measuring}
G\'{a}bor~J Székely, Maria~L Rizzo, and Nail~K Bakirov.
\newblock Measuring and testing dependence by correlation of distances.
\newblock 2007.

\bibitem[Wang et~al.(2012)Wang, Jiang, Wang, Zhou, and Tu]{wang2012unsupervised}
Bo~Wang, Jiayan Jiang, Wei Wang, Zhi-Hua Zhou, and Zhuowen Tu.
\newblock Unsupervised metric fusion by cross diffusion.
\newblock In \emph{2012 IEEE Conference on Computer Vision and Pattern Recognition}, pages 2997--3004, 2012.
\newblock \doi{10.1109/CVPR.2012.6248029}.

\bibitem[Wang et~al.(2014)Wang, Mezlini, Demir, Fiume, Tu, Brudno, Haibe-Kains, and Goldenberg]{wang2014similarity}
Bo~Wang, Aziz~M Mezlini, Feyyaz Demir, Marc Fiume, Zhuowen Tu, Michael Brudno, Benjamin Haibe-Kains, and Anna Goldenberg.
\newblock Similarity network fusion for aggregating data types on a genomic scale.
\newblock \emph{Nature Methods}, 11\penalty0 (3):\penalty0 333--337, 2014.

\end{thebibliography}






\end{document}